\documentclass[10pt,conference, letterpaper]{IEEEtran}
\IEEEoverridecommandlockouts

\usepackage[T1]{fontenc}
\usepackage[utf8x]{inputenc}
\usepackage{amsmath,amssymb,amsfonts}
\usepackage{mathtools}
\usepackage{graphicx}
\usepackage{booktabs}
\usepackage{xspace}
\usepackage{multirow}
\usepackage{xcolor}
\usepackage[colorlinks=true,urlcolor=teal,
            linkcolor=teal,citecolor=teal]{hyperref}
\usepackage{cleveref}
\usepackage{balance}
\usepackage{enumitem}
\usepackage{url}
\usepackage{csquotes}
\usepackage{glossaries-extra}
\usepackage{tikz}
\usepackage{orcidlink}

\usetikzlibrary{patterns,decorations.pathmorphing,calc,positioning}
\setabbreviationstyle[acronym]{long-short}
\glssetcategoryattribute{acronym}{nohyperfirst}{true}

\newacronym{ftqc}{FTQC}{Fault Tolerant Quantum Computers}
\newacronym{icc}{ICC}{Intraclass Correlation Coefficient}
\newacronym{nisq}{NISQ}{Noisy Intermediate Scale Quantum}
\newacronym{pec}{PEC}{Probabilistic Error Cancellation}
\newacronym{qec}{QEC}{Quantum Error Correction}
\newacronym{qem}{QEM}{Quantum Error Mitigation}
\newacronym{qft}{QFT}{Quantum Fourier Transformation}
\newacronym{qtc}{QTC}{Quantum Trotter Circuit}
\newacronym{zne}{ZNE}{Zero-Noise Extrapolation}

\usepackage{censor}
\newboolean{anonymous}
\setboolean{anonymous}{false}
\ifbool{anonymous}{}{}
\ifbool{anonymous}{}{\renewcommand{\blackout}[1]{#1}}
\ifbool{anonymous}{\newcommand{\genemail}[2]{\href{mailto:xxx.xxx@xx.xx}{\blackout{#2}}}}{\newcommand{\genemail}[2]{\href{#1}{#2}}}

\newcommand{\repro}{\href{https://github.com/lfd/qce26_artefactual_improvements_zne}{reproduction package\xspace}}
\newcommand{\zenodo}{\href{https://doi.org/10.5281/zenodo.21540565}{Zenodo archive\xspace}}

\newcommand{\etal}{\emph{et al.}\xspace}

\usepackage[backend=biber,style=ieee,maxnames=3,citetracker=true,citestyle=ieee-comp,defernumbers=true]{biblatex}

\newbibmacro*{title+link}[1]{%
  \iffieldundef{doi}
    {%
      \iffieldundef{url}
        {#1}%
        {\href{\thefield{url}}{\textcolor{teal}{#1}}}%
    }
    {\href{https://doi.org/\thefield{doi}}%
       {\textcolor{teal}{#1}}}%
}

\DeclareFieldFormat
  [online,article,inbook,incollection,inproceedings,patent,thesis,unpublished,misc]
  {title}{%
    \usebibmacro{title+link}{#1}%
  }

\renewbibmacro*{doi+eprint+url}{}
\renewbibmacro*{url+urldate}{}

\AtEveryBibitem{
	\clearfield{pages}
	\clearfield{isbn}
	\clearfield{eprint}
	\clearfield{issn}
	\clearfield{abstract}
	\clearfield{volume}
	\clearfield{number}
	\clearfield{month}
}

\renewbibmacro{in:}{}
\AtBeginBibliography{\footnotesize}

\begin{document}

\bstctlcite{BSTcontrol}

\title{Benchmarking Error Mitigation: Artefactual Improvements in Zero-Noise Extrapolation}

\author{
    \IEEEauthorblockN{\blackout{Dominik Köster\orcidlink{0009-0001-0230-3123}}}
    \IEEEauthorblockA{
        \blackout{\textit{Technical University of}} \\
        \blackout{\textit{Applied Science Regensburg}} \\
        \blackout{Regensburg, Germany} \\
        \genemail{mailto:dominik.koester@othr.de}{dominik.koester@othr.de}
    }
    \and
    \IEEEauthorblockN{\blackout{Wolfgang Mauerer\orcidlink{0000-0002-9765-8313}}}
    \IEEEauthorblockA{
        \blackout{\textit{Technical University of}} \\
        \blackout{\textit{Applied Science Regensburg}} \\
        \blackout{\textit{Siemens AG, Foundational Technology}} \\
        \blackout{Regensburg/Munich, Germany} \\
        \genemail{mailto:wolfgang.mauerer@othr.de}{wolfgang.mauerer@othr.de}
    }
}

\maketitle
 
\begin{abstract}
Reliable benchmarking of \gls{qem} requires distinguishing genuine improvements from artefacts of the post-processing arithmetic. In this paper, we expose a failure mode in Richardson \gls{zne}, a widely used technique routinely (and often implicitly) relied upon in benchmarks and experiments. When noise amplification operates beyond usable signals~--~a regime that is quickly reached on current hardware for non-trivial circuits~--~we show that the extrapolation no longer reflects the underlying physics, but collapses into a fixed rescaling of a single noisy measurement, producing a bogus \emph{apparent improvement} that is independent of noise amplification. This poses a rarely considered threat to the validity of many empirical evaluations in quantum computing.

Measurements on real hardware (IQM Euro-Q-Exa) confirm this collapse with ordinary folding alone: as circuit depth erodes the signal, the reported estimate decouples from the truth and overshoots the ideal by up to 21\%. We further introduce a matched-cost \enquote{garbage-folding} negative control that carries no usable signal yet reports a \emph{larger} apparent improvement than genuine folding~--~showing that the magnitude of an improvement is not evidence of its correctness~--~alongside a zero-cost check flagging the artefact from data a benchmark already holds. We distil both into a short reporting checklist for \gls{zne} benchmarks.
\end{abstract}
\vspace{-1em}

\section{Introduction}
\label{sec:intro}
Hardware noise systematically distorts the outcomes of \gls{nisq} algorithms~\cite{greiwe_effects_2023,thelen_approximating_2024}, and is a crucial obstacle to any practical or industrial use~\cite{carbonelli_challenges_2024}. \gls{zne} has become a widely used \gls{qem} technique to counter this~\cite{temme_error_2017,li_efficient_2017,endo_practical_2018,cai_quantum_2023}, as it is hardware-agnostic and requires little computational overhead. The Richardson variant~\cite{temme_error_2017,endo_practical_2018, giurgica-tiron_digital_2020, krebsbach_optimization_2022} fits a polynomial to expectation values at different noise amplification levels $\lambda_k$ and extrapolates to the zero-noise limit $\lambda=0$. The extrapolated value $\hat{E}(0)$ is reported as the benchmark outcome, and the magnitude of the improvement $\hat{E}(0)-E(\lambda_1)$ is taken as evidence of successful mitigation. Benchmarking error mitigation therefore hinges on a deceptively simple question: is a reported improvement real, or merely an artefact of the post-processing pipeline?

All of this rests on the assumption that the measured expectation values still decay predictably with $\lambda$. If the true signal decays faster than the polynomial model assumes, the extrapolation can yield an unphysical improvement that stems not from the intended noise amplification mechanism, but from a mathematical artefact: a destroyed signal that the extrapolation arithmetic rescales into an apparent improvement that is often over-corrected or possibly unphysical. Govia~\etal~\cite{govia_bounding_2025} described a related effect for \gls{pec}~--~an apparent improvement produced by the post-processing arithmetic rather than the intended mechanism~--~which they term the \emph{horoscope effect}. We adopt this name for its Richardson \gls{zne} analogue, give it a closed form, and demonstrate it on hardware with a matched-cost negative control, extending our broader study of statistical artefacts in \gls{qem} benchmarks~\cite{koester_claim_2026} in the benchmarking-by-validation spirit of testing what a pipeline delivers~\cite{russo_testing_2023} under reproducible protocols~\cite{mauerer_repro_2022}.

For a credible benchmark, apparent improvements must be attributable to the intended noise amplification mechanism rather than an extrapolation arithmetic. We provide four contributions:
\begin{itemize}
  \item a closed-form characterisation of \emph{when and why} Richardson \gls{zne} degenerates to deterministic rescaling, pinpointing the regime in which a benchmark's reported improvement becomes meaningless;
  \item a signal-retention taxonomy that lets a benchmark decide, from its own per-$\lambda$ data, which regime a circuit is in, across simulation and the IQM Euro-Q-Exa hardware;
  \item a matched-cost \enquote{garbage-folding} negative control that quantifies how much apparent improvement a benchmark can manufacture from signal destruction alone, demonstrated on hardware against genuine folding as the positive reference;
  \item a zero-cost negative-probability check and a short reporting checklist that flag the artefact from data a benchmark already holds.
\end{itemize}

\section{Background: Richardson Extrapolation}
\label{sec:background}
\gls{zne}~\cite{li_efficient_2017,temme_error_2017} estimates a noise-free expectation value $\hat{E}(0)$ from measurements at $K$ amplified noise levels $\lambda_1<\dots<\lambda_K$. Richardson extrapolation fits a degree-$(K{-}1)$ polynomial through the data; the zero-noise estimate is the linear combination as 
\begin{equation}
  \hat{E}(0)=\sum_{k=1}^{K}c_k\,E(\lambda_k),\qquad \sum_{k=1}^{K}c_k=1,
  \label{eq:richardson}
\end{equation}
whose Lagrange coefficients $c_k$ depend only on the chosen scale factors~\cite{cai_quantum_2023}. Digital \gls{zne} can realise $\lambda_k>1$ by gate-level unitary folding $G\mapsto G(G^\dagger G)^k$, which leaves the ideal circuit invariant while multiplying its error rate~\cite{giurgica-tiron_digital_2020}. We quantify the improvement by the recovery ratio $\rho = \frac{\hat{E}(0)-E(\lambda_1)}{E_\text{ideal}-E(\lambda_1)}$, the fraction of the gap between the raw value $E(\lambda_1)$ and the ideal value $E_\text{ideal}$ that the extrapolation closes: $\rho=0$ means no improvement over the raw value, $\rho=1$ a perfect recovery of the ideal value, and $\rho>1$ an unphysical overshoot beyond it.

\paragraph{Variance amplification}
As it holds that $\mathrm{Var}(\hat{E}) = \sum_{k=1}^K c_k^2\, \mathrm{Var}(E(\lambda_k))$, the noise-amplification factor is bounded by $\sum_{k=1}^K |c_k|$,  and can be determined from 
scale factors
alone~\cite{cai_quantum_2023,krebsbach_optimization_2022}. For the
widely used set $\{1,3,5\}$~\cite{giurgica-tiron_digital_2020,majumdar_best_2023}
one obtains $\sum|c_k|=3.5$ with $c_1=\tfrac{15}{8}$, $c_2=-\tfrac{5}{4}$,
$c_3=\tfrac{3}{8}$; the tightly spaced set $\{1,1.1,1.25,1.5\}$~\cite{kandala_error_2019} yields
$\sum|c_k|=681$, a $194\times$ larger variance bound from scale-factor choice
alone.

\paragraph{The signal-retention assumption}
Equation \eqref{eq:richardson} is informative only if $E(\lambda_k)$ carries signal beyond statistical noise. Under ideal depolarising noise, a probability observable decays towards the floor $f = 1/2^N$ for $N$ qubits. A parity observable $\langle Z^\otimes N \rangle$ decays towards $f=0$ and can even become negative under noise, which is not uncommon on real hardware. If $E(\lambda_k)$ approaches the floor, the extrapolation can yield an unphysical improvement, which can already occur at $\lambda=3$ for circuits with non-trivial depth on current hardware (see Euro-Q-Exa experiment in Figure \ref{fig:horoscope}).

\section{Degeneration to Deterministic Rescaling}
\label{sec:degeneration}
Suppose that for every amplified scale factor, the expectation value collapses to the floor, $E(\lambda_k)\approx f$ for $k>1$. Substituting into
\eqref{eq:richardson} and using $\sum_{k=1}^K c_k=1$, we find
\begin{equation}
  \hat{E}(0)=c_1E(\lambda_1)+\underbrace{\Big(\textstyle\sum_{k>1}c_k\Big)}_{=\,1-c_1}f
            =c_1E(\lambda_1)+(1-c_1)\,f.
  \label{eq:deterministic}
\end{equation}
For $\{1,3,5\}$, this leads to a closed-form function of the \emph{single} noisy value $E(\lambda_1)$: $\hat{E}(0)=\tfrac{15}{8}E(\lambda_1)-\tfrac{7}{8}f$. For high qubit counts, $f$ collapses to near zero, and three observations follow: (1) the extrapolation is independent of the noise amplification method and yields the same $\hat{E}(0)$; (2) since $c_1>1$, any $E(\lambda_1)$ is rescaled upward, resulting in an apparent improvement that can overshoot the ideal value; and (3) the apparent improvement persists even for a non-functional amplification method, making it indistinguishable from a genuine improvement without further diagnostics. The argument is not specific to Richardson or scale factors $\{1,3,5\}$: Any linear extrapolator with $c_1>1$ collapses to the same single-point rescaling once $E(\lambda{>}1)$ reaches the floor.

\begin{figure}[htbp]
  \centering
  \includegraphics{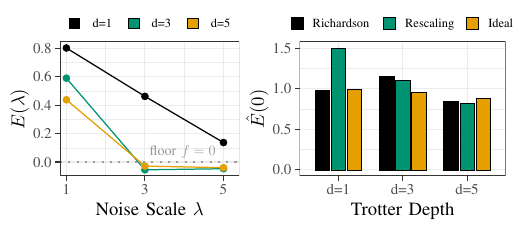}
  \caption{Richardson extrapolation on Euro-Q-Exa. Left: The
    amplified parity values $E(\lambda)$ of a 4-qubit \gls{qtc} collapse to the
    floor $f=0$ as the Trotter depth $d$ grows. Right: At each depth, the Richardson
    estimate $\hat{E}(0)$, the rescaling prediction $\tfrac{15}{8}E(\lambda_1)$,
    and $E_\text{ideal}$. Once the signal reaches the floor ($d{=}3,5$),
    $\hat{E}(0)$ tracks the rescaling and is decoupled from the ideal value;
    at $d{=}1$, where the signal is retained, it instead recovers the ideal.}
  \label{fig:rescaling}
\end{figure}

\paragraph{Empirical confirmation on hardware}
\Cref{eq:deterministic} is not merely a worst case: it is reached by ordinary, faithful folding on real hardware. We measure the parity observable $\langle Z^{\otimes 4}\rangle$ (floor $f=0$) of a 4-qubit \gls{qtc} on a connected, low-error chain of the 54-qubit IQM Euro-Q-Exa machine, using the scale factors $\{1,3,5\}$ folding throughout (protocol in \Cref{sec:hardware}), while increasing the Trotter depth $d$. The left-hand panel of \Cref{fig:rescaling} shows the amplified values $E(\lambda{>}1)$ collapsing to the floor as $d$ grows. The right-hand panel compares, at each depth, the Richardson estimate $\hat{E}(0)$ with the rescaling prediction $\tfrac{15}{8}E(\lambda_1)$ and with $E_\text{ideal}$. At $d{=}1$ the signal is retained ($E(\lambda_3){=}0.46$), the two predictions differ markedly, and Richardson recovers the ideal value. At $d{=}3,5$ the amplified values have reached the floor and $\hat{E}(0)$ tracks the pure rescaling to within $5\%$, decoupled from the truth: at $d{=}3$ it overshoots the ideal by $21\%$ ($\hat{E}{=}1.16$ against $E_\text{ideal}{=}0.96$), at $d{=}5$ it merely happens to fall below. The number of negative extrapolated per-state estimates (see \Cref{sec:diagnostic}) grows in lockstep, from $3/16$ to $7/16$.

\section{A Signal-Retention Taxonomy}
\label{sec:taxonomy}
\begin{figure}[htbp]
  \includegraphics{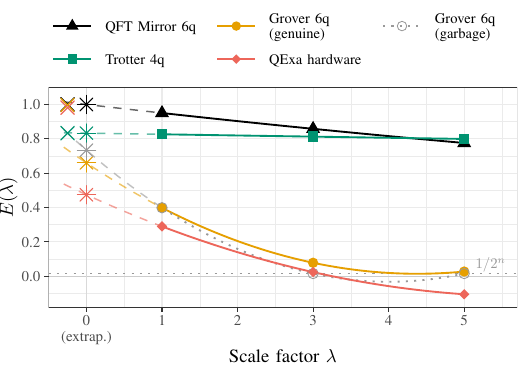}\vspace*{-1em}
  \caption{Richardson extrapolation across signal-retention regimes at
    $\lambda\in\{1,3,5\}$. Solid: Lagrange polynomial through $E(\lambda_k)$;
    dashed: extrapolation to $\hat{E}(0)$ (stars); crosses mark $E_\text{ideal}$.
    Grey dotted: garbage folding on Grover. The Euro-Q-Exa parity signal turns
    negative at $\lambda=5$ (signal destroyed). Horizontal dotted: floor
    $1/2^6$ of the 6-qubit circuits.}
  \label{fig:horoscope}
\end{figure}

\Cref{fig:horoscope} shows $E(\lambda)$ for four circuits under simulated
depolarising noise ($p_\text{2q}=10^{-3}$), and for Euro-Q-Exa. Three qualitatively distinct regimes emerge:
\begin{enumerate}[noitemsep,topsep=2pt,leftmargin=1.2em]
  \item \textbf{Signal retained.} The 6-qubit \gls{qft} mirror and the 4-qubit
    \gls{qtc} decay slowly and stay well above the floor at $\lambda=5$;
    Richardson recovers near-ideal values ($\rho=0.99$ and $\rho=1.00$).
  \item \textbf{Signal decayed.} The 6-qubit Grover circuit reaches
    $E(\lambda_3)=0.078$ and $E(\lambda_5)=0.027$, both near the floor
    $f=1/64$, and recovery degrades to $\rho=0.44$.
  \item \textbf{Signal destroyed.} On Euro-Q-Exa, the parity signal is negative
    at $\lambda=5$, satisfying the precondition of \Cref{eq:deterministic}: any
    improvement on this device is partly an artefact of the rescaling arithmetic.
\end{enumerate}
Recovery degrades \emph{progressively} as $E(\lambda{>}1)$ approaches the floor:
the fit loses informative anchors and $\rho$ drops accordingly.

\section{Garbage-Folding Falsification}
\label{sec:garbage}
\subsection{Design}
Apart from hardware runs, all circuits are simulated under ideal depolarising noise, for which Richardson extrapolation is exact~\cite{cai_quantum_2023}. To probe regimes with stronger noise than in this model~\cite{maschek_make_2025}, and to check if improvement in the destroyed regime depends on the amplification method, we replace genuine folding with a \enquote{garbage folding}: Inserted folds do not reduce to the identity, $U(U^\dagger U)\neq U$, yet match genuine folding in gate count and residual error rate. This maximises signal destruction at $\lambda>1$, the exact precondition of \Cref{eq:deterministic}, and acts as a \emph{negative control}. Matched in cost but carrying no usable signal, it isolates whether a larger reported improvement reflects better mitigation or merely stronger signal destruction.

\subsection{Simulation}
The grey dotted line in \Cref{fig:horoscope} shows garbage folding on the Grover circuit outperforming genuine folding ($\rho\approx0.56$ versus $0.44$). This illustrates a general failure mode: whenever the actual noise at $\lambda>1$ exceeds the polynomial model~--~through non-Markovian effects, coherent errors, or simply higher-than-expected gate errors~--~Richardson overestimates the noise-free value, an artefact of the coefficient arithmetic rather than the amplification. Applied to the signal-retained \gls{qft} and \gls{qtc}, the same garbage folding drives $\rho$ to the wildly unphysical $16$ and $125$, which is an extreme but unambiguous instance of the same effect.

\subsection{Two routes to unphysical extrapolation}
\label{sec:spectrum}
The degeneration of \Cref{sec:degeneration} is one of two distinct ways Richardson \gls{zne} can turn unphysical, and the two must not be conflated. \emph{(1) Ill-conditioning:} closely spaced scale factors carry a large coefficient sum $\sum|c_i|$, which by the variance bound of \Cref{sec:background} amplifies statistical noise and residual model error into an overshoot even when every $E(\lambda)$ retains signal. This route is well understood and is precisely why the community favours the widely spaced set $\{1,3,5\}$~\cite{giurgica-tiron_digital_2020,majumdar_best_2023,cai_quantum_2023}, whose bias and variance are bounded in closed form by~\cite{mohammadipour_direct_2025}. Recent work characterises the resulting finite-shot help--harm boundary~\cite{alfaro_finite-shot_2026}. On the Grover circuit, as $\sum|c_i|$ grows by two orders of magnitude, genuine recovery degrades from faithful at $\{1,3,5\}$ ($\rho_\text{gen}=0.44$, $\sum|c_i|=3.5$) to a sixfold overshoot for the closely spaced Kandala set $\{1,1.1,1.25,1.5\}$~\cite{kandala_error_2019} ($\rho_\text{gen}=6.0$, $\sum|c_i|=681$). \emph{(2) Signal destruction:} the route studied here is orthogonal, governed by the proximity of $E(\lambda{>}1)$ to the floor, not by $\sum|c_i|$, and therefore strikes even the variance-optimal $\{1,3,5\}$ as soon as depth or hardware noise destroys the signal (see \Cref{fig:rescaling}).

\subsection{Hardware}
\label{sec:hardware}
We repeat the falsification on the 54-qubit IQM Euro-Q-Exa machine (via the \href{https://munich-quantum-software-stack.github.io/MQSS-Interfaces/qiskit/index.html}{MQSS Qiskit adapter}). Faithful benchmarking requires that the designed circuit is executed under identical conditions every time. Concretely, the linear \gls{qtc} is mapped onto a directly-connected, low-error qubit chain (qubits $8\text{--}11$ from the calibration), transpiled at \texttt{optimization\_level}${=}0$ onto the native gate set, with the \texttt{no\_modify} flag set to prevent register re-routing. Pinning a connected chain (an earlier non-adjacent register incurred routing overhead and worse values) makes genuine and garbage folding differ only in the inserted fold copies, sharing an identical two-qubit (\textsc{cz}) count at every $\lambda$. Each circuit is sampled with $4096$ shots, so the per-$\lambda$ binomial standard error on the parity stays below $0.02$, which is far smaller than the signal collapse and the overshoot we report, which therefore cannot be attributed to shot noise. 

On this calibration $E(\lambda)$ decays from $0.77$ to $0.41$ to $0.12$ at $\lambda\in\{1,3,5\}$, against $E_\text{ideal}=0.98$. Averaged over $15$ matched repetitions, genuine folding already overshoots slightly ($\rho_\text{gen}=0.99\pm0.13$) because the signal sits near the floor at $\lambda{>}1$, whereas garbage folding overshoots far beyond it ($\hat E_\text{garb}=1.38\pm0.04$, $\rho_\text{garb}=2.77\pm0.25$) while driving $10$ of $16$ basis-state estimates negative. This is a larger apparent improvement than the genuine one, yet entirely unphysical. The overshoot is deterministic rather than slow drift: a blocked-versus-interleaved acquisition bounds the inter-circuit bias below $1.4\times10^{-3}$ and the Allan deviation follows the $1/\sqrt{m}$ white-noise law, so only longer averaging~--~not the folding~--~reduces it (details in our \repro{}). Two controls isolate the cause. An identity fold (extra native gate pairs composing to the identity, matched in count to garbage folding) neither overshoots nor triggers the artefact ($\rho_\text{id}=0.74\pm0.09$), so the ghost improvement comes from signal-\emph{destroying} folds, not from added gates. And where signal is retained, genuine folding genuinely helps~--~lowering the mean-squared error against the raw $\lambda{=}1$ baseline by more than an order of magnitude~--~so the failure is specific to the artefact regime, not to \gls{zne} itself.

\section{A Zero-Cost Negative-Probability Diagnostic}
\label{sec:diagnostic}
Constraining the extrapolation to the physical range of the observable suppresses unphysical scalar estimates~\cite{miranskyy_improving_2026}, yet an artefactual estimate can easily fall within that range~--~as seen for $d{=}5$ in \Cref{fig:rescaling}, the pure rescaling merely happens to fall below the ideal value. A better indicator of artefact improvement is the full probability distribution of the results, that is, the results of applying Richardson to every basis state: For each state $s$, $\hat{P}(s)=\sum_k c_kP_s(\lambda_k)$. A valid distribution requires $\hat{P}(s)\ge0$ for all $s$. Small violations are expected even for genuine folding. The negative coefficient $c_2=-5/4$ pushes a few estimates marginally below zero under statistical noise, but a large violation is a fingerprint of the degeneration regime: Once $P_s(\lambda_{3,5})$ have collapsed to the floor, $c_2$ drives a substantial share of the per-state estimates well below zero. On the 6-qubit Grover circuit at $p_\text{2q}=6.4\times10^{-4}$, genuine folding leaves all states valid ($0/64$ negative, $\min_s\hat{P}(s)=+0.0005$) while garbage folding pushes nearly half below zero ($29/64$ negative, $\min_s\hat{P}(s)=-0.039$). The check requires no additional computational cost, as a benchmark already holds the per-state counts at each $\lambda$. Reporting the negative fraction and $\min_s\hat{P}(s)$ alongside $\hat{E}(0)$ turns a hidden artefact into a visible check.

This diagnostic is not a simulation artefact; on hardware it discriminates perfectly. Across $405$ Euro-Q-Exa runs spanning all three regimes ($45$ repetitions each of \{genuine, identity, garbage\} folding $\times$ \{retained, decayed, destroyed\}), we summarise the violation by the \emph{negative-probability weight} $W_\text{neg}=\sum_{s:\hat{P}(s)<0}|\hat{P}(s)|$. Unlike the recovery ratio $\rho$, which requires the ideal value and diverges as $E_\text{ideal}\!\to\!E(\lambda_1)$ (the source of the unphysical $\rho=16,125$ in \Cref{sec:garbage}), $W_\text{neg}$ needs no ground truth and stays bounded, making it the more robust benchmarking statistic. Valid folding (genuine, identity) stays an order of magnitude below the boundary in every regime ($W_\text{neg}=0.02$--$0.04$), whereas garbage folding sits far above it ($0.74\pm0.06$); the two never overlap (largest valid $0.06$ versus smallest garbage $0.64$; AUC$=1.0$). The garbage value is moreover regime-invariant (coefficient of variation ${<}8\%$): once the signal is destroyed, \Cref{eq:deterministic} fixes the per-state estimate to a single rescaling, so $W_\text{neg}$ saturates against a ceiling set by the scale factors alone. It is therefore a binary validity flag, available from the per-$\lambda$ counts, that flags the artefact however convincing $\hat{E}(0)$ looks.

\section{Implications and Conclusion}
\label{sec:conclusion}
We have shown that Richardson extrapolation degenerates to deterministic rescaling whenever noise amplification destroys the signal at $\lambda>1$: Any folding can manufacture an apparent \enquote{improvement} larger than the genuinely achieved advantage, which constitutes a horoscope effect. We confirmed the effect on hardware with genuine $\{1,3,5\}$ folding alone. While hardware evidence is limited to one device and circuit family, simulation broadens circuit coverage but assumes ideal depolarising noise. As \Cref{eq:deterministic} depends only on $\sum_k c_k=1$ and the floor $f$, we expect the failure mode to generalise, with onset depth and noise level being device-specific. Crucially, this signal-destruction route is distinct from the familiar ill-conditioning of closely spaced scale factors: controlled by proximity to the floor rather than by $\sum|c_i|$, it leaves even the variance-optimal $\{1,3,5\}$ fully exposed.

From this we distil a reporting checklist that any Richardson \gls{zne} benchmark can adopt at no additional computational cost:
\begin{enumerate}[label=(\arabic*),noitemsep,topsep=2pt,leftmargin=1.8em]
  \item report whether $E(\lambda)$ still retains signal at $\lambda>1$ or has reached the observable floor;
  \item report the per-basis-state negative-probability weight $W_\text{neg}$ alongside $\hat{E}(0)$, not the scalar estimate alone;
  \item report improvement relative to the physically attainable maximum, and flag overshoots; where the ideal value is unknown, bound it with a classically simulable surrogate and flag any $\hat{E}(0)$ exceeding the bound.
\end{enumerate}
A failure of any item marks the reported improvement as an artefact rather than genuine mitigation. Paired with the matched-cost garbage-folding negative control~--~which a benchmark suite can run as a routine sanity check~--~these items separate what a \gls{zne} pipeline claims to measure from what it actually measures.

\begin{small}
\textbf{Data availability} Code and data to generate the complete paper are available in our \repro{} and a long-term \zenodo{}. HW calibration snapshots and logs allow reproduction without machine access.

\textbf{Acknowledgments}
The authors gratefully acknowledge the use of the quantum system Euro-Q-Exa, co-funded by the EuroHPC JU, BMFTR (grant 13N16690), and the Bavarian State Ministry of Science and the Arts, operated by the Leibniz Supercomputing Centre (LRZ) in Garching, Germany, for providing the computational resources for this work. We acknowledge partial support by the German Research Foundation (DFG), grant MA 9739/1-1, and the High-Tech Agenda of the Free State of Bavaria. We also acknowledge partial support by the European Regional Development Fund (ERDF) and by the Free State of Bavaria as part of the project AIM-SMEs (Grant No. 2506-014-3.2), co-funded by the European Union. 
\end{small}

\printbibliography
 
\end{document}